\documentclass[sigconf]{acmart}
\renewcommand\footnotetextcopyrightpermission[1]{} 

\AtBeginDocument{%
  \providecommand\BibTeX{{%
    \normalfont B\kern-0.5em{\scshape i\kern-0.25em b}\kern-0.8em\TeX}}}

\setcopyright{none}

\begin{document}

\title{Low-Rank Representations Towards Classification Problem of Complex Networks}

\author{
Murat Çelik }
\email{b21827263@cs.hacettepe.edu.tr}
\affiliation{%
      \institution{Department of Computer Engineering, \\ Hacettepe University}
  \city{Ankara}
  \country{Turkey}
  \postcode{06800}
}

\author{
Ali Baran Taşdemir}
\email{alibaran@tasdemir.us}
\affiliation{%
  \institution{Department of Computer Engineering, \\ Hacettepe University}
  \city{Ankara}
  \country{Turkey}
  \postcode{06800}
}

\author{
Lale Özkahya}
\email{ozkahya@cs.hacettepe.edu.tr}
\affiliation{%
  \institution{Department of Computer Engineering, \\ Hacettepe University}
  \city{Ankara}
  \country{Turkey}
  \postcode{06800}
}


\begin{abstract}
Complex networks representing social interactions, brain activities, molecular structures have been studied widely to be able to understand and predict their characteristics as graphs. Models and algorithms for these networks are used in real-life applications, such as search engines, and recommender systems. In general, such networks are modelled by constructing a low-dimensional Euclidean embedding of the vertices of the network, where proximity of the vertices in the Euclidean space hints the likelihood of an edge (link).  In this work, we study the performance of such low-rank representations of real-life networks on a network classification problem. 
\end{abstract}





\maketitle
\pagestyle{plain}

\section{Introduction}
Complex networks representing social interactions, brain activities, molecular structures have been studied widely to be able to understand and predict their characteristics as graphs. Building good models for complex networks, considering the role of social networks in understanding modern human interaction, is very important \cite{barabasi1999emergence, chakrabarti2006graph, watts1998collective}. In recent years, obtaining low-dimensional embeddings of graph nodes in Euclidean space has been frequently studied in machine learning applications,and in this way, thanks to the geometry of the embedding, it has been tried to preserve the structural properties of the graph \cite{hamilton2017inductive}. Specifically, the node embedding method represents each node in the n-node graph as a vector of $x\in \mathbb{R}^k$ and $k<<n$. In this way, the embedding method is used in many subjects such as clustering, classification, link prediction, together with machine learning \cite{hamilton2017inductive}. In this sense, the probability of an edge/link between two nodes can be deduced from the geometric proximity of the vectors representing them.

Two of these embedding methods that are widely used are TSVD (Truncated Singular Value Decomposition) and PCA (Logistic Principal Component Analysis) methods such as decomposition of the adjacency matrix of the given network (or graph) \cite{ahmed2013distributed}. In recent years, the method of learning graph embedding using deep neural networks has been frequently used \cite{cao2016deep,grover2016node2vec, perozzi2014deepwalk, tang2015line}. The question these methods are trying to solve is to get a given large networks to be represented by matrices of smaller dimensions than the original, while preserving its structural properties. One of the still unanswered questions is how small of a matrix this representation can be done while preserving the structural information of the network \cite{seshadhri2020impossibility, chanpuriya2020node, loukas2019graph, garg2020generalization}. For example, the frequent occurrence of triangles is an important characteristic in social networks such as Facebook, and the extent to which this can be preserved by graph embedding techniques such as TSVD and PCA has been researched \cite{seshadhri2020impossibility, chanpuriya2020node}, \cite{ stolman2022classic}, binary node classification question, the class labels of the nodes were applied as graph characteristics. 

The main contribution of this work is to examine the performance of the commonly used TSVD and LPCA graph embedding techniques in the multiclass classification problem of the original graph. The graph classification question is a problem that has been studied in many different fields, and in some cases it has difficulties such as trying to extract class information without sufficient network information. Bonner et al.~\cite{bonner2016deep} developed a learning method for graph classification by using the topology features of the network using deep neural networks. One of the comprehensive studies in terms of the variety of random graph models was done by Rossi and Ahmed~\cite{rossi2019complex}. In this study, 9 network classes with distinctly different structures, including network groups constructed with random network models, are included in the classification question. It has been observed that the reconstructed networks to be represented at low dimensions by graph embedding methods largely preserve their class properties. As a result of the experiments, it was seen that the LPCA method gave better performance than the TSVD method. It has been observed that random networks need much larger sizes than real world networks in order to be successfully represented.

\begin{table*}[htbp]
\caption{The average values for each network group of the attributes used in the classification process.}
\label{tab:features}
\resizebox{\textwidth}{!}{%
\begin{tabular}{|c|c|c|c|c|c|c|c|c|c|}
\hline
\textbf{Graph Classes} &
  \textbf{Number of Node} &
  \textbf{Number of Edge} &
  \textbf{Density} &
  \textbf{Max. Degree} &
  \textbf{Avg. Degree} &
  \textbf{Max. k-core} &
  \multicolumn{1}{c|}{\textbf{\begin{tabular}[c]{@{}c@{}}Avg. Clustering \\ Coefficient\end{tabular}}} &
  \textbf{Number of Triangles} &
  \multicolumn{1}{c|}{\textbf{\begin{tabular}[c]{@{}c@{}}Avg. Eigenvector \\ Centrality\end{tabular}}} \\ \hline
Barabasi-Albert & 1000 & 35409  & 0.0709 & 278  & 71  & 40  & 0.127 & 767320  & 0.0247 \\ \hline
Biological      & 2715 & 27895  & 0.0098 & 339  & 20  & 37  & 0.256 & 272918  & 0.0096 \\ \hline
Brain           & 487  & 12561  & 0.2358 & 245  & 48  & 35  & 0.520 & 398615  & 0.0610 \\ \hline
Chung-Lu        & 5201 & 21437  & 0.0045 & 167  & 8   & 16  & 0.072 & 9449    & 0.0105 \\ \hline
Economic        & 2283 & 100090 & 0.1574 & 863  & 120 & 111 & 0.437 & 5518546 & 0.0255 \\ \hline
Enzymes         & 60   & 94     & 0.0948 & 6    & 4   & 3   & 0.287 & 21      & 0.0989 \\ \hline
Erdos-Renyi     & 1000 & 37583  & 0.0752 & 99   & 75  & 62  & 0.075 & 410869  & 0.0310 \\ \hline
Facebook        & 6314 & 231472 & 0.0138 & 987  & 76  & 64  & 0.258 & 1910524 & 0.0086 \\ \hline
Retweet         & 5586 & 6396   & 0.0005 & 1972 & 2   & 4   & 0.013 & 236     & 0.0055 \\ \hline
\end{tabular}
}
\end{table*}

\section{Method}

In the classification process, different real network groups such as biology, economy and social networks, as well as different network groups produced by random methods, constitute the classes of the classification process. The performance of TSVD and LPCA methods is compared to how much the representation of the original networks given by the graph embedding methods by the graphs in lower dimensions preserves the class feature.Classification is done by using the graph features, which are known to play an important role in graphs. 
\subsection{Graph Embedding Methods} 
\paragraph{TSVD :}TSVD (Truncated Singular Value Decomposition) is a spectral method that approximates the adjacency matrix by using SVD (Singular Value Decomposition). The TSVD method is used to reconstruct the adjacency matrix.  For an adjacency matrix $A \in \left\{ 0, 1 \right\}^{N \times N}$, let $Z \in \mathbb{R}^{N \times k}$ be the orthonormal matrix. The columns of the orthonormal matrix $Z$ include the eigenvectors of the adjacency matrix corresponding to the k largest magnitude eigenvalues. And we define $W$ as diagonal with entries corresponding to the top k eigenvalues. Then we calculate TSVD embeddings by $X = Zs(W)\sqrt{|W|}$ and $Y = Z\sqrt{|W|}$. The $s()$ function represents the sign function. To compute the reconstructed adjacency matrix we compute $\sigma(XY^T)$ where $\sigma$ is a clipping function between 0 and 1.
\paragraph{LPCA : } 
We use another method based on the LPCA (Logistic PCA) to approximate the adjacency matrix. With a logistic loss function, we are trying to minimize the error between $\sigma(XY^T)$ (approximation) and A (adjacency matrix). For a given $A \in \left\{ 0, 1 \right\}^{N \times N}$, and $X, Y \in \mathbb{R}^{N \times k}$, we use the loss function:

$$
L = \sum_{i=1}^{N} \sum_{j=1}^{N}-\log l \left( \tilde{A}_{i,j}\left[ XY^T \right]_{i,j} \right)
$$

where $l(x) = (1 + e^{-x})^{-1}$ is logistic function and $\tilde{A}$ is the shifted adjacency matrix with -1's in place of 0's ($\tilde{A} = 2A - 1$). We initialize X and Y with random values between $\left[ -1, 1 \right]$. And we find X and Y embeddings that minimize the loss function by using SciPy \cite{jones2001scipy} implementation of the "L-BFGS-B" \cite{byrd1995limited} algorithm with 100 iterations. And we reconstruct the approximated adjacency matrix by using $\sigma(XY^T)$.

In both of these graph embedding methods, the {\it rank} of operation is the $k$ value seen in the size of the obtained matrices.

\subsection{Graph Attributes and Models} 

The features used are density, maximum degree, average degree, maximum k-core, average clustering coefficient, total number of triangles, average eigenvector centrality, which are also seen in Table~\ref{tab:features}. Below are the definitions of these graph attributes. A graph is defined as $G=(V,E)$ with the node set $V$ and the edge set $E$. The parameters $n=|V|$ and $m=|E|$ will represent the number of elements of these sets. The number of neighboring nodes connected to each node is the {\it degree} of that node, and the maximum, minimum and average (over $n$) values defined on the graphs are among the attributes that play an important role in areas such as coloring and matching problems. The value of {\it graph density}, one of the most important graph parameters affecting the algorithmic runtime, is calculated as $\frac{m}{{n\choose 2}}$. In this context, if a graph contains $n^2$ edges, it means that it is dense, and sparse otherwise.

A graph's clustering coefficient measures the clustering tendency of the nodes in the graph~\cite{watts1998collective}. {\it Eigenvector centrality} measures the effect of one node on other nodes. The eigenvector centrality of a node is the value of the digit corresponding to that node in the eigenvector of the largest eigenvalue of the adjacency matrix. If a node has a high degree, that node will have a high eigenvector centrality. Another important network parameter is the $k$-core of the diagram~\cite{sariyuce2017parallel,sariyuceliu2019analysis}. The k-core of a graph is the number of nodes in the largest possible subgraph that includes at least $k$ neighbors of each node in it. 

\subsection{Classification Process} 

Random Forest (RF), Stochastic Gradient Descent (SGD), K-Nearest Neighbor (KNN) and Support Vector Machines (SVM) Classifiers are used as Machine Learning model. In the training phase, F1 scores are obtained by stratified, 10-fold cross validation, and optimization study is carried out with the help of these scores.\textit{F1 score} is a statistical measure calculated by taking the harmonic average of precision and recall values. These scores are observed as 0.87, 0.77, 0.86, and 0.89 for Random Forest (RF), Stochastic Gradient Descent (SGD), K-Nearest Neighbor (KNN), and Support Vector Machines (SVM), respectively. A second F1 score is obtained on the optimized models by determining 20\% of the training data set as the test data set. Support Vector Machines (SVM) with an F1 score of 0.92 among the results are observed as the most suitable model for classification. For this reason, the performances of graph embedding methods are compared over the F1 score results obtained with SVM in the following sections. 

\section{Experimental Results} 
For the experiments, a total of 198 graphs from 9 different classes are studied~\cite{networkrep}. These network groups and their number of elements: Biology Networks (bio, 30), Brain Networks (bn, 10), Economy Networks (econ, 16), Retweet Networks (rt, 29), Facebook Networks (fb, 30), Enzymes Networks ( enz, 30), Barab\'asi–Albert Networks (ba, 13), Chung-Lu Networks (cl, 24), Erd{\H o}s–R\'enyi Networks (er, 16),the attributes of the networks are given in Table-\ref{tab:features} as the mean value. The last three of these network groups have been created with the named random network models and are used to represent different network structures.

\begin{table}[h]
  \centering
  \caption{Average values of F1-scores found with the SVM classifier over all network groups when using different ranks of graph embedding methods.}
  \label{tab:svmresults}
  \begin{tabular}{cccc}
    \hline
\multicolumn{2}{c}{TSVD} & \multicolumn{2}{c}{LPCA}\\
    Rank & F1 Score & Rank & F1 Score \\
    \hline
16 & 0.48 & 5 & 0.51 \\
32 & 0.52 & 16 & 0.82 \\
64 & 0.70 & 32 & 0.92 \\
128 & 0.81 & 64 & 0.97 \\
    \hline
  \end{tabular}
\end{table} 

The main distinction between random network models is that the probability of the edge added between two nodes differs according to the model. In the BA model \cite{barabasi1999emergence}, as a node is added at each step, each new edge is determined in direct proportion to the degrees of pre-existing nodes. In the ER model \cite{paul1959random}, the number of node pairs is ${n\choose 2}$, and each pair turns into an edge with a fixed probability $p$. The CL model \cite{chung2002average} generalizes the ER model to fit the expectation of a given node degree (number of neighbors) distribution with unequal edge probabilities. 
\subsection{Classification Accuracy} 
Since the SVM provides the highest F1 score during the training phase, the values of this classifier are taken as a reference in terms of performance during the testing phase. When the mean F1 score across all networks are compared, the values seen in Table \ref{tab:svmresults} are found to be 0.70 and 0.81 for TSVD (for grade = 64 and 128), LPCA (for grade = 64 and 128) 0.92 and 0.97, respectively. From this point of view, LPCA provides higher accuracy for the same rank than TSVD.

\begin{figure}[h]
    \centering
    \includegraphics[width=\linewidth]{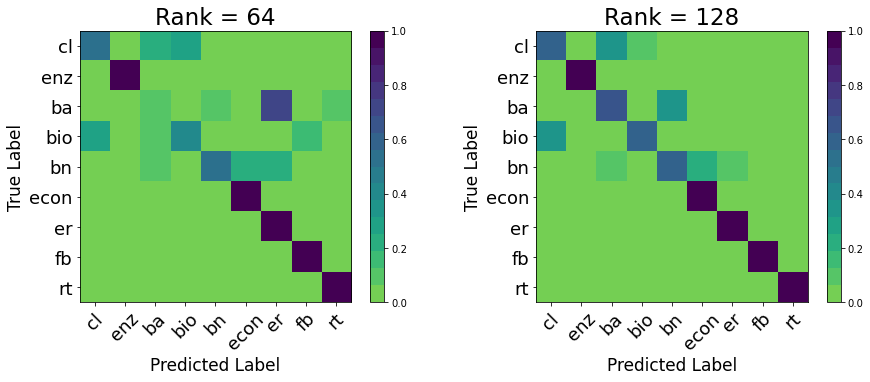}
    \caption{Confusion matrix obtained by TSVD method and evaluating classification accuracy over all network groups.}
    \label{fig:confmatrixTSVD}
\end{figure}

\begin{figure}[h]
    \centering
    \includegraphics[width=\linewidth]{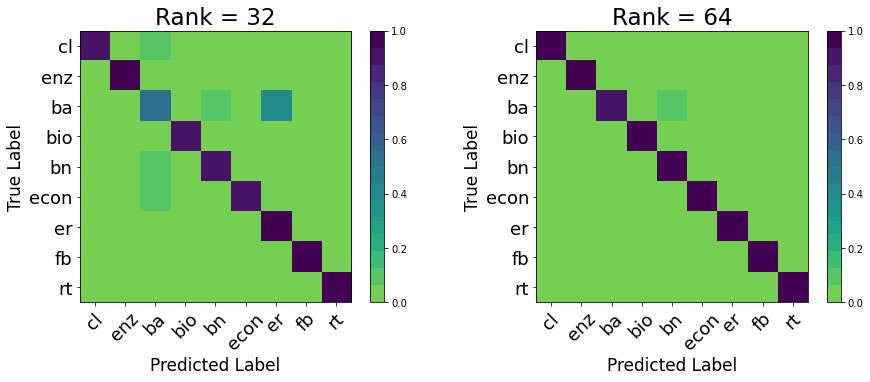}
     \caption{Confusion matrix obtained by LPCA method and evaluating classification accuracy over all network groups.}
    \label{fig:confmatrixLPCA}
\end{figure}
When the confusion matrices seen in Figure \ref{fig:confmatrixTSVD} and \ref{fig:confmatrixLPCA} are compared, it is also seen that the classification performance of LPCA is superior to TSVD. In addition, classification performance is low in networks such as BA, CL. In particular, the performance of the LPCA for the BA group is weaker than for the other groups. In this sense, on the classification question, it is concluded that these methods could not achieve the same success as for real-life networks in random networks.  

\begin{table}[hbt!]
\caption{Values of the (minimum) factorization dimension size obtained using the LPCA method, providing successful classification with an F1 score of 92\% for the tested network groups.}
\label{tab:exactdim}
\resizebox{\linewidth}{!}{%
\begin{tabular}{lccc}
\hline
\textbf{Graph Classes} & \textbf{Number of Node} & \textbf{Average Degree} & \textbf{Rank} \\ \hline
Biological      & 2715 & 19.58  & 5  \\
Enzymes         & 60   & 3.47   & 5  \\
Retweet         & 5585 & 2.34   & 5  \\
Brain           & 487  & 47.50  & 16 \\
Chung-Lu        & 5201 & 7.96   & 16 \\
Economic        & 2283 & 120.02 & 16 \\
Erdos-Renyi     & 1000 & 75.17  & 32 \\
Facebook        & 6314 & 76.10  & 32 \\
Barabasi-Albert & 1000 & 70.82  & 64 \\
\hline
\end{tabular}
}
\end{table}

Table \ref{tab:exactdim} lists the values of the (minimum) factorization dimension size obtained using the LPCA method, which provides all networks in the group to be classified with an F1 score of 92\% for the network groups tested. Looking at these values, it is observed that the LPCA method gives successful results in graph embedding with a very small rank value for Biology, Enzymes and Retweet networks. As seen in Table \ref{tab:features}, these networks are networks with low density and average degree. Generally, in random networks (BA, CL, ER), higher rank is required to reach the same level of precision.

{\it Calculation Time :} 
For comparison purposes, one network each from the Economy and BA random network group is selected and the computational speed of the LPCA and TSVD methods is compared in Table \ref{tab:runtime}. The computation time for TSVD is found to be much shorter than for LPCA. It is predicted that this time spent for LPCA will increase further if the number of iterations increases. The same difference is seen for large networks other than the two example networks given here, making TSVD preferable in terms of computation time. 
\begin{table}[h]
  \centering
  \caption{Comparison of the computation times (in seconds) of the two methods on econ-beause (n=0.5K, m=44K) and ba\_1k\_150k (n=1K, m=150K) networks, for purposes of example.}
  \label{tab:runtime}
  \begin{tabular}{ccccc}
    \hline
&
\multicolumn{2}{c}{econ-beause} & \multicolumn{2}{c}{ba\_1k\_150k}\\
    
Rank &   TSVD & LPCA & TSVD & LPCA \\
    \hline
 16 & 0.40 & 3.54 & 1.75 & 15.98 \\
 32 & 0.44 & 3.86 & 1.65 & 14.33 \\
 64 & 0.36 & 2.60 &  1.70 & 15.19 \\
    \hline
  \end{tabular}
\end{table} 
\newpage
\section{Conclusions} 
In this study, it is seen that the networks reconstructed to be represented at low dimensions by graph embedding methods largely preserved their class properties.On the other hand, in similar studies, it is seen that some features, such as the triangle distribution, which are important in the analysis of networks, are not sufficiently preserved by the methods presented here. In this respect, the presented methods need improvement to preserve different graph properties in the reconstruction process.


\bibliographystyle{ACM-Reference-Format}
\bibliography{Main_ENG}

\appendix

\end{document}